\documentclass[sigconf,authorversion]{acmart}
\AtBeginDocument{%
  \providecommand\BibTeX{{%
    \normalfont B\kern-0.5em{\scshape i\kern-0.25em b}\kern-0.8em\TeX}}}

\copyrightyear{2021} 
\acmYear{2021} 
\setcopyright{acmcopyright}\acmConference[AH2021]{12th Augmented Human International Conference}{May 27--28, 2021}{Geneva, Switzerland}
\acmBooktitle{12th Augmented Human International Conference (AH2021), May 27--28, 2021, Geneva, Switzerland}
\acmPrice{15.00}
\acmDOI{10.1145/3460881.3460937}
\acmISBN{978-1-4503-9030-9/21/05}
\begin{document}

\title{NaviChoker: Augmenting Pressure Sensation via Pneumatic Actuator}

\author{Shogo Yoshida}
\email{s2020047@jaist.ac.jp}
\orcid{1234-5678-9012}
\affiliation{%
  \institution{Japan Advanced Institute of Science and Technology}
  \state{Ishikawa}
  \country{Japan}
  \postcode{923-1292}
}

\author{Haoran Xie}
\email{xie@jaist.ac.jp}
\orcid{1234-5678-9012}
\affiliation{%
  \institution{Japan Advanced Institute of Science and Technology}
  \state{Ishikawa}
  \country{Japan}
  \postcode{923-1292}
}

\author{Kazunori Miyata}
\email{miyata@jaist.ac.jp}
\orcid{1234-5678-9012}
\affiliation{%
  \institution{Japan Advanced Institute of Science and Technology}
  \state{Ishikawa}
  \country{Japan}
  \postcode{923-1292}
}

\renewcommand{\shortauthors}{S. Yoshida, et al.}

\begin{abstract}
  Many technologies have been developed in recent years to present audiovisual information in new ways, but developing an information presentation interface to convey tactile information is still a challenge. We propose a tactile device using wearable technology that is an all-around pressure presentation system using pneumatic actuators. Specifically, we develop a system in which a choker equipped with a pneumatic actuator is worn around the neck, that applies pressure in any direction to indicate to the user the direction in which to walk and also when to start and stop walking. In this paper, we describe the construction of the device, evaluation experiments, our assessment of the prototype, and future plans for the device.
\end{abstract}

\ccsdesc[500]{Human-centered computing~Haptic devices}

\keywords{pneumatic acutator, pressure sensation, haptic, choker, interaction}


\maketitle

\begin{figure}[ht]
  \centering
  \includegraphics[scale=0.4]{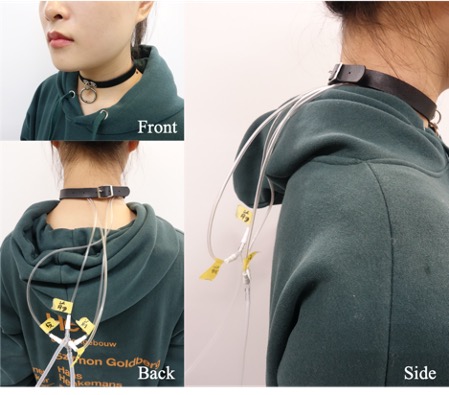}
  \caption{NaviChoker using pneumatic actuators.}
\end{figure}

\section{Introduction}

Interaction and sensing using various stimuli in virtual worlds is an important part of applications of virtual and augmented reality. Haptic interactions using the sense of touch to detect force can be applied to numerous scenarios in the field of human-computer interaction (HCI) \cite{thermo, vib}. The emerging technologies of haptic devices can expand the user experience by conveying various types of information, enhancing entertainment, education, and our daily lives. In this study, we propose a wearable device to be used in the real world that delivers pressure sensations.

Various devices have been developed in HCI that deliver a pressure sensation. LevioPole uses eight propellers to effect translation and rotation in any direction in the air \cite{pole}. The Force Jacket uses a pneumatic actuator inside the jacket to apply various pressures on the user \cite{force}. However, these devices usually require heavy setup or a large working space, which may make them less comfortable and easy to use as wearable devices.

To solve these issues, we propose a lightweight and independent choker-type device using mini-pneumatic actuators, NaviChoker, as shown in Figure 1. NaviChoker utilizes four small pneumatic actuators mounted on the user’s neck. The proposed device can be used in various applications, such as security notifications from smartphones, map navigation, and haptic interaction in virtual reality (VR) applications. The proposed device may be especially useful in noisy environments as an alternative to screen displays or voice guidance. In addition, this device can be used as a fashion item as a choker.

\section{Related Works}
In the field of human augmentation, the wearable devices have been proposed extensively to augment human physical capabilities \cite{xie19}. However, it is challenging to augment user sensation via wearable devices. xClothes uses retractable structures to augment human thermo-sensation \cite{cloth20}. The Interactive SPA Skin \cite{soft} uses multiple small pneumatic actuators to present various haptic sensations to the arm to simulate tactile sensations. In this work, we use pneumatic actuators for augmenting pressure sensation.

There are several Wearable devices proposed for sensation augmentation. For visual sensation, the head-mounted devices were proposed with anamorphosis projection \cite{yoshida2021}. For haptic sensation, a neck-mounted haptic device was proposed to dynamically modulate vibrations to lead the user to reach the destination based on haptic information \cite{ele}. ActiveBelt \cite{belt} presents intuitive directional information to the user by activating a vibrator attached to the belt. LevioPole can give walking directions to the user, but as the device is large, it cannot be used in confined spaces \cite{pole}. The Force Jacket can provide a pressure sensation to the body but can only be used in an enclosed space such as a room, since it uses a large air compressor to supply air to the actuator \cite{force}. In thermal or electrical stimulation as haptic modes of presentation \cite{song}, the device needs to be worn closely against the user’s neck to present equal stimulation to the user. In soft robotics, the pneumatic actuators are usually proposed for robotic grippers as bending function \cite{qi2021bpactuators}.  In this work, the pneumatic actuator can provide haptic presentation regardless of how tightly the device is worn.

\section{NaviChoker}
The device we propose in this study can apply pressure to the entire circumference of the user’s neck using a pneumatic actuator, as shown in Figure 2. This device can apply equal pressure to the entire circumference of the user’s neck without the choker sticking to the neck, due to the actuator’s mechanism of inflation by air pressure. In addition, since this device can be worn as a choker, it can continuously present directional information to the user.

The proposed device uses actuators attached to the inside of the choker to apply pressure from any direction around the user’s neck. In addition, this system can apply a variety of haptic interactions to the user due to the individual and continuous actuator operations. The advantage of the proposed device is that the user can feel pressure from the actuators without close contact with the device itself.

\begin{figure}[t]
  \centering
  \includegraphics[scale=0.4]{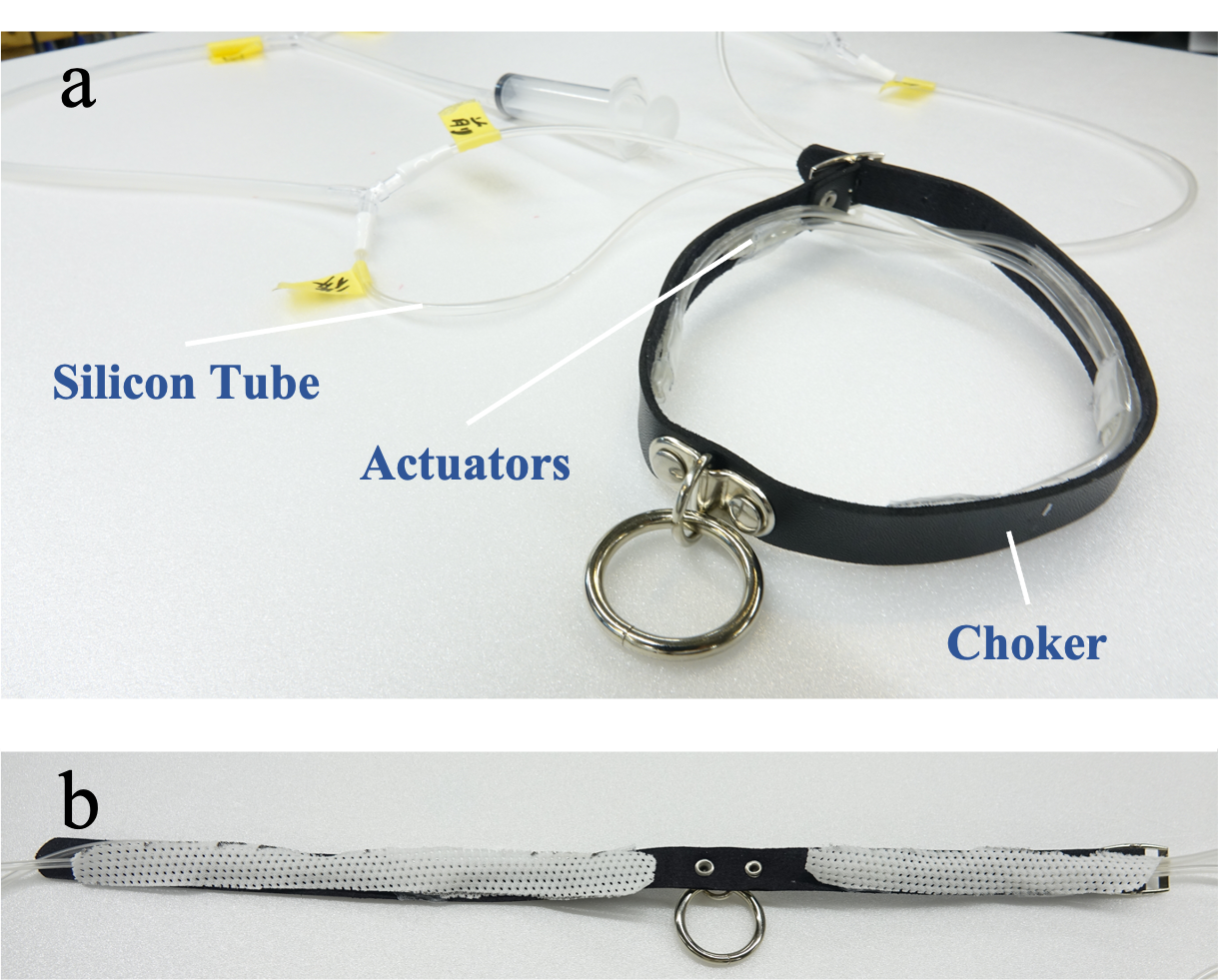}
  \caption{Prototype of the proposed NaviChoker (a); a cloth cover attached inside the device (b).}
\end{figure}

\subsection{Pneumatic Actuator}

Figure 3 shows the pneumatic actuator we developed for our proposed device. The actuator uses polyethylene film of 0.08 mm thickness, which is cut into $2.0 \times 3.0$ cm squares, and a silicon tube. We used a soldering iron to seal the polyethylene film.

The tip of the silicon tube was bent to prevent air from escaping. A 1.0 $mm^2$hole was made in the side. Similarly, a 1.0 $mm^2$ hole was drilled in the center of polyethylene film, and the holes were glued together.

\begin{figure}[t]
  \centering
  \includegraphics[scale=0.5]{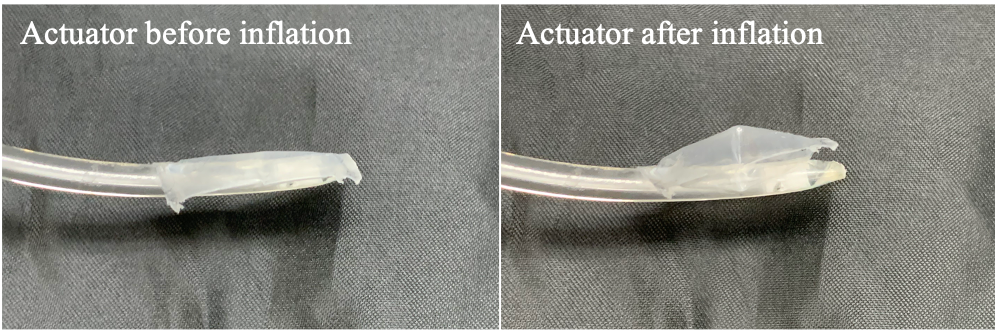}
  \caption{Actuation performance of the fabricated pneumatic actuator.}
\end{figure}

\begin{figure}[t]
  \centering
  \includegraphics[scale=0.45]{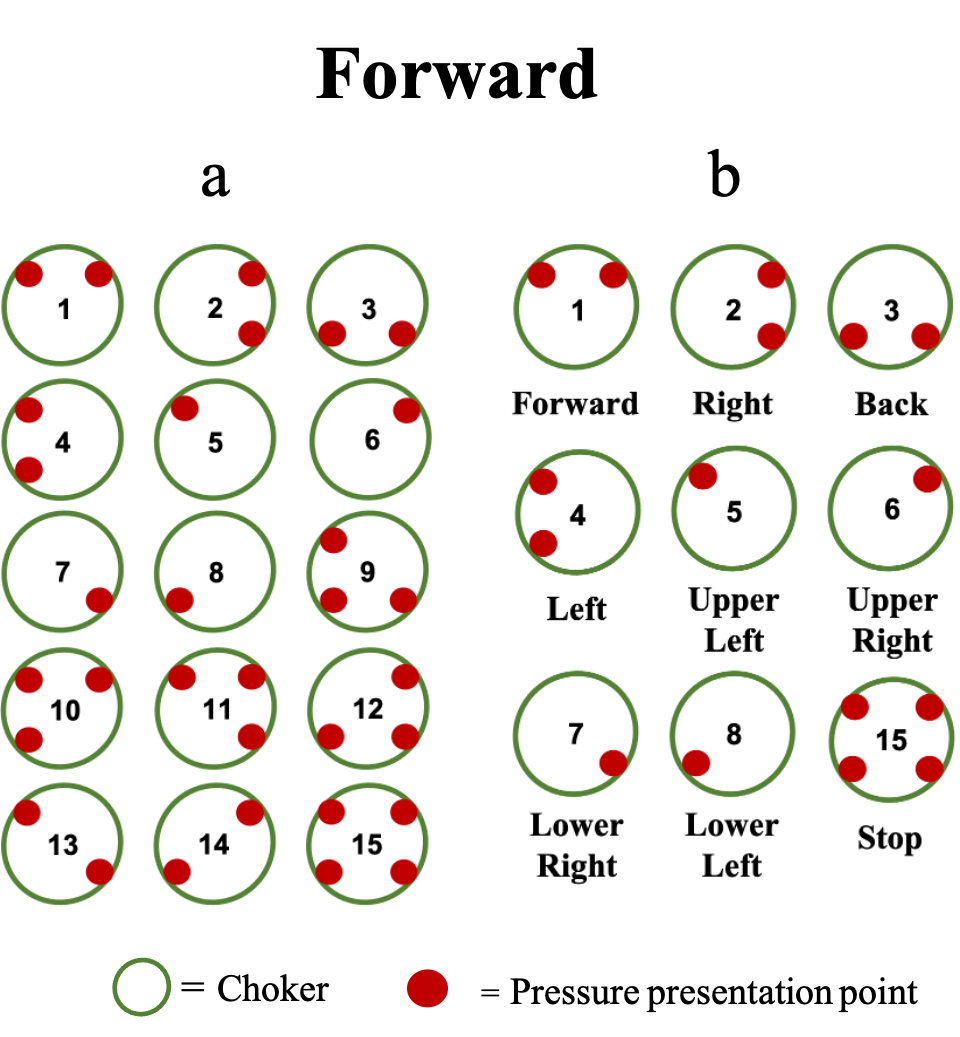}
  \caption{All patterns of pressure presentation used in preliminary experiments (a); pressure pattern by presentation direction used in the experiment (b).}
\end{figure}

\subsection{Prototype Design}

Figure 2(a) shows a NaviChoker prototype made with a pneumatic actuator. This device can present pressure in eight direction through the four actuators attached in equal intervals around the user’s neck. For example, if the front two actuators are inflated, the user can attend to the forward direction. In addition, we covered the actuators with a soft cloth to improve wearing comfort and to stabilize the actuators, as shown in Figure 2(b). The actuators are intended to operate automatically in connection with an electronic devices, but for the prototype, the actuators were manually operated using syringes to test that the sensation is sufficient.

\subsection{Preliminary Study}

We conducted a preliminary study to explore the relationship between the directions of pressure presentation and the actual stimulus directions based on user sensation. We invited seven participants (1 female, 6 males, ages 24–32 years old) in our preliminary study. The parameters explored include the number of actuators and the actuator locations of pressure presentation.

In this study, we presented 15 different pressure patterns to users to test the perceived directions indicated, as shown in Figure 4(a). From this study, appropriate combinations of actuator motions for eight directions was selected (Figure 4(b)).

\section{User Study}

To verify the usefulness of the proposed NaviChoker device, we set a navigation task as a case study to help users to reach a designated destination with multiple interactions. We conducted a questionnaire survey after the user test. Although the proposed device can be used for various applications, this work focuses on giving walking directions by signaling users in certain directions.

Figure 5 shows the setting of the experiment and Figure 6 shows the route to the destination. We conducted a walking experiment to complete the evaluation of the proposed device. There were nine participants (2 females, 7 males, ages 24–32 years old). We adopted a “Wizard of Oz” approach to control the actuator. When a user approached a turn, the experimenter activated a pneumatic actuator using a syringe to indicate the direction to turn towards. In addition, users had noise-canceling earphones attached so as to walk relying only on the pressure sensations.

In this experiment, we also applied pressure to indicate the start and end of walking. The start pressure is the first direction indicated at the start of the experiment, with the end indicated as the pattern numbered 15 in Figure 4(b). 

After the experiment, we conducted a questionnaire survey to confirm the device’s usability, how it feels to wear it, and the pressure presentation of this system. Users responded to the following questions with ratings out of five stars:

\begin{itemize}
\item[1] Did you clearly feel the pressure presented by the actuator?
\item[2] Did you clearly feel the differences between directional presentations?
\item[3] Do you feel any discomfort due to the pressure presentations?
\end{itemize}

\begin{figure}[ht]
  \centering
  \includegraphics[scale=0.55]{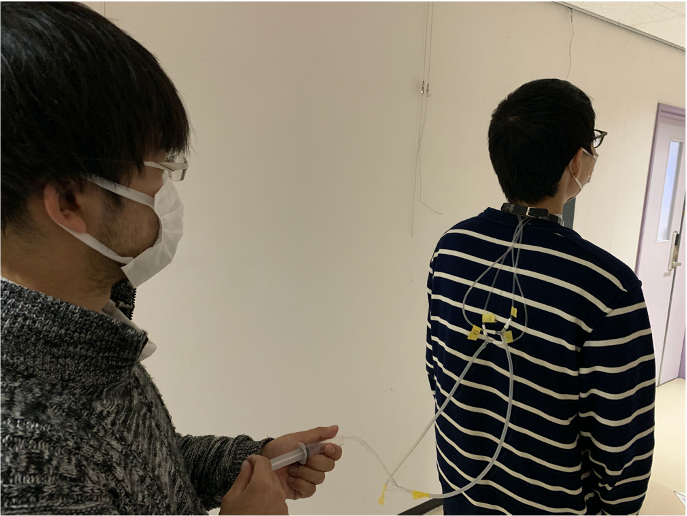}
  \caption{A participant in our user study.}
\end{figure}

\begin{figure}[ht]
  \centering
  \includegraphics[scale=0.33]{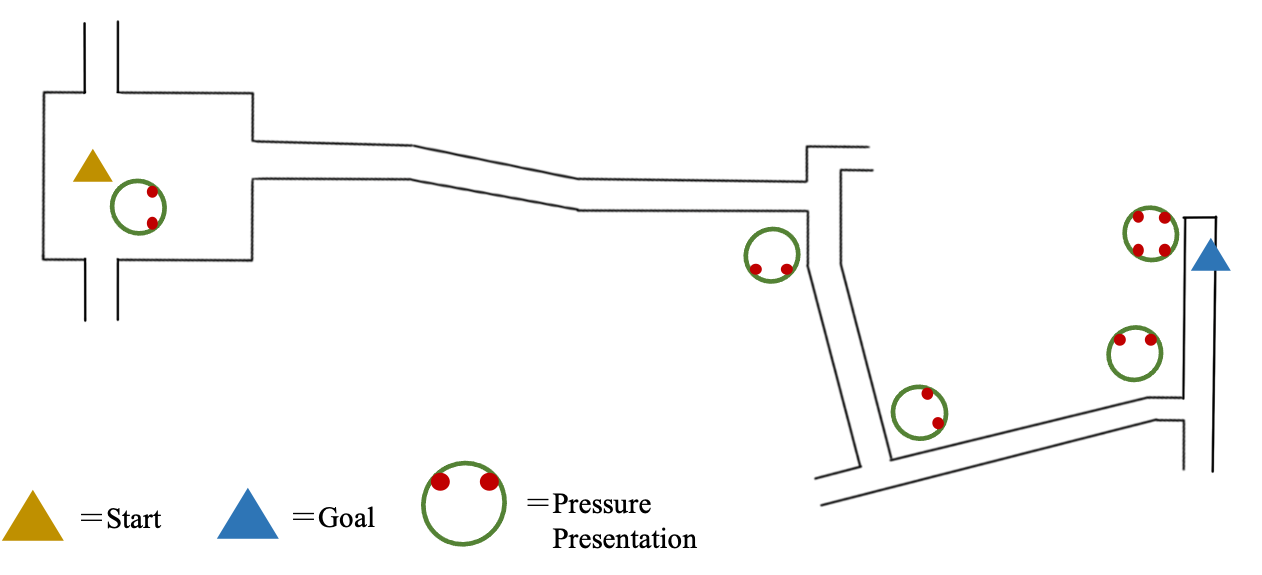}
  \caption{Walking route and pressure presentation locations in the user study.}
\end{figure}

\section{Results}

Eight users were able to reach the target location by means of the presented pressure alone. NaviChoker was able to correct the direction of the users by simply presenting pressure to users who were facing in a different direction from the target direction before the start of the experiment. In addition, users were able to recognize the pressure that starts and ends walking.

This indicates that the pneumatic actuators can be used as a navigation system. Navigation systems that use a screen display and voice guidance have some drawbacks such as continuing walking direction while on the phone and difficulty using it in noisy areas. We suggest that the proposed device overcomes these problems. However, one subject was not able to reach the destination. The reason for this was that the user perceived the presented pressure as an indication to go backward just before the end of the experiment. The position and strength of the felt pressure may therefore differ among subjects, so we need to further investigate the appropriate pressure strength and presentation position for different users to improve the device so that equal pressure can be felt among different users.
Figure 7 shows the results from the questionnaire. The items survey differences in directional presentation, the degree of pressure presentation, and the evaluation of discomfort when wearing the device. The results show that the pressure presentation was highly rated, as all of the subjects felt the presented pressure. In addition, many of the subjects felt the differences in the direction indications.

\begin{figure}[ht]
  \centering
  \includegraphics[scale=0.4]{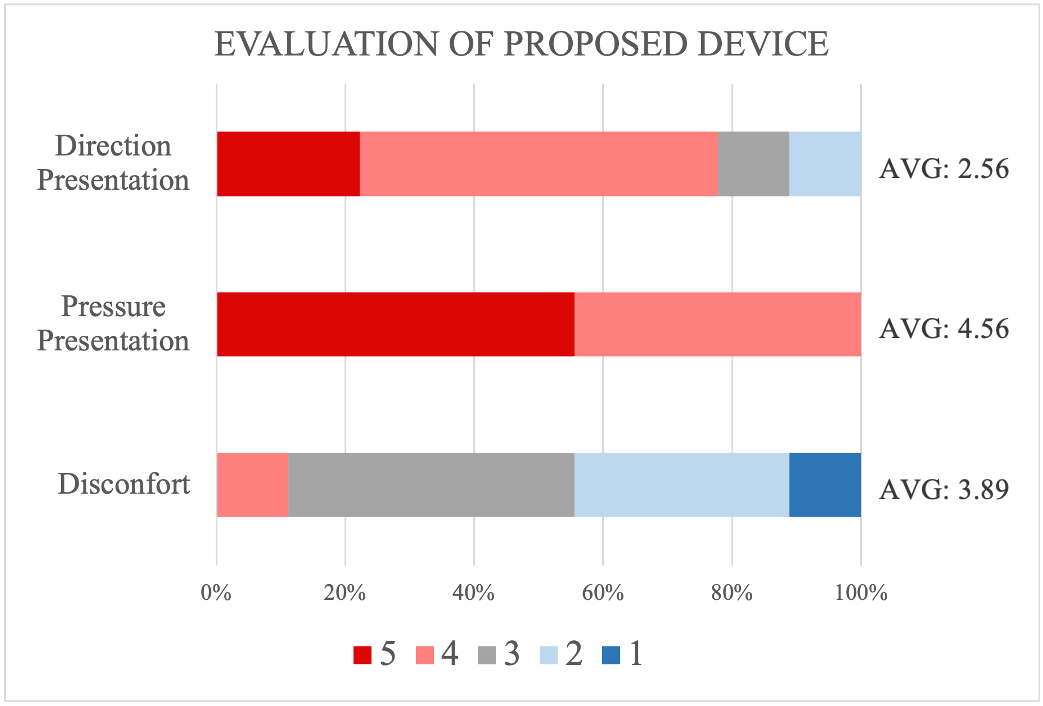}
  \caption{Evaluation results of our user study.}
\end{figure}

However, some participants’ feedback indicate that improvements to the device are needed. Some of the participants did not feel much difference in pressure direction. The reason may be that the actuator could not reach the neck due to individuals’ neck being too thin, or that the actuator impacted parts of the neck where haptic sensation was difficult to feel. This problem could be solved by increasing the extent of the actuator’s inflation and the area of installation around the neck. Four users felt that wearing the device was uncomfortable. This may be caused by an unfamiliar feel of the choker due to poor device design. As a solution, a soft cloth may be placed between the neck and the device to reduce the feeling of the mechanism to the user. Also, the actuator mechanism needs to be downsized and the silicon tube embedded in the choker.

\section{Conclusion}

In this paper, we proposed a choker-style haptic device, NaviChoker, for augmenting pressure sensations. The proposed device can provide omnidirectional pressure presentation with pneumatic actuators. We confirmed differences in user perception of the pressure presented and the usefulness of the actuator in navigation tasks. In particular, the pneumatic actuator was shown to be sufficiently useful for presenting walking directions because users could clearly perceive the applied pressure.

NaviChoker can be improved for various pressure presentations. More detailed pressure presentation can be achieved by increasing the number of actuators. The harder actuator materials may provide clearer haptic presentation. The proposed device can be used for various applications besides navigation , such as notifications, timers, and entertainment purposes. An actuator can change the intensity and pattern of the haptic presentation according to the application or the sender of the message in the notification function. In addition, haptic presentation similar to Morse code or Braille can be used as a communication tool by applying various combinations of actuator motions. 

For future work, automating the pneumatic actuators should provide significant improvements. In this study, the actuator was operated manually by a “Wizard of Oz” method. However, this approach does not allow for rapid changes in directional presentation. With automated control, it may be possible to accurately control the strength, speed, and number of times the actuators inflate. In addition, more complex and flexible motion patterns may be conveyed by increasing the number of actuators. Improvements in the actuator may also help achieve achieve continuous motions. It should also be possible to present haptic sensations to the entire circumference of the wearer’s neck by placing multiple pneumatic actuators. To embed an automatic pneumatic actuator into a choker, we need to design a compact pneumatic circuit (e.g., Bubble \cite{bubble}). A small device may be developed, for example, by using a micro-pump (micro ring pomp RP-Q1), a small solenoid valve (KOGANEI Petit Valve), and a micro-computer (Arduino). We can reduce the burden on the user by making the device lighter and smaller than the current prototype.

Potential applications such as navigation and entertainment require interaction between the user and the device. The proposed device can achieve this by attaching a GPS sensor or IR camera. Different device designs may serve for different applications. Therefore, we need to investigate and develop linkages between the proposed device and external sensors. The actuator used in this study was made from two rectangular pieces of plastic film stacked on top of each other that inflates from the center. Therefore, haptic sensations may differ if the device is designed with different modes of inflation. It should be useful to investigate the differences in users’ haptic sensations due to differences in inflation and various haptic presentations. In addition, the worn comfort of the device needs to be improved based on our survey results.


\begin{acks}
We greatly thank Ping Zhang for photographing of the prototype image. This work was supported by JSPS KAKENHI grant JP20K19845.
\end{acks}

\bibliographystyle{ACM-Reference-Format}
\bibliography{sample-base}

\end{document}